\documentclass{book}
\newdimen\footheight
\makeatletter
\newcommand\vpt   {\edef\f@size{\@vpt}\rm}
\newcommand\vipt  {\edef\f@size{\@vipt}\rm}
\newcommand\viipt {\edef\f@size{\@viipt}\rm}
\newcommand\viiipt{\edef\f@size{\@viiipt}\rm}
\newcommand\ixpt  {\edef\f@size{\@ixpt}\rm}
\newcommand\xpt   {\edef\f@size{\@xpt}\rm}
\newcommand\xipt  {\edef\f@size{\@xipt}\rm}
\newcommand\xiipt {\edef\f@size{\@xiipt}\rm}
\newcommand\xivpt {\edef\f@size{\@xivpt}\rm}
\newcommand\xviipt{\edef\f@size{\@xviipt}\rm}
\newcommand\xxpt  {\edef\f@size{\@xxpt}\rm}
\newcommand\xxvpt {\edef\f@size{\@xxvpt}\rm}
\makeatother
\usepackage{rmh,makeidx}
\usepackage{amsmath,amsfonts,graphicx,subfigure,url}

\usepackage{euscript}
\newcommand{\Eu}[1]{\ensuremath{\EuScript{#1}}}
\newcommand{\eps}{\varepsilon}
\newcommand{\R}{\ensuremath{\mathbb{R}}}
\newcommand{\cost}{\ensuremath{\textsf{cost}}}
\newcommand{\dist}{\ensuremath{\textsf{dist}}}

\begin{document}
\setcounter{chapter}{49}

\runhds{J.M. Phillips}
{Chapter 49: Coresets and Sketches}

\thispagestyle{plain}

\vspace{-1pc}

\noindent\hskip-3.72pc{\hvbxxiv 49\hskip 27pt CORESETS AND SKETCHES}

\vspace{1pc}

\noindent{\hvxiv Jeff M. Phillips}

\vspace{4pc}

\textheight=50pc

\Bnn{INTRODUCTION}

\noindent
Geometric data summarization has become an essential tool in both geometric approximation algorithms and where geometry intersects with big data problems.  In linear or near-linear time large data sets can be compressed into a summary, and then more intricate algorithms can be run on the summaries whose results approximate those of the full data set.
Coresets and sketches are the two most important classes of these summaries.

A {\trmbitx coreset} is a reduced data set which can be used as proxy for the full data set; the same algorithm can be run on the coreset as the full data set, and the result on the coreset approximates that on the full data set.
It is often required or desired that the coreset is a subset of the original data set, but in some cases this is relaxed.
A \emph{weighted coreset} is one where each point is assigned a weight, perhaps different than it had in the original set.
A \emph{weak coreset} associated with a set of queries is one where the error guarantee holds for a query which (nearly) optimizes some criteria, but not necessarily all queries; a \emph{strong coreset} provides error guarantees for all queries.

A {\trmbitx sketch} is a compressed mapping of the full data set onto a data structure which is easy to update with new or changed data, and allows certain queries whose results approximate queries on the full data set.
A \emph{linear sketch} is one where the mapping is a linear function of each data point, thus making it easy for data to be added, subtracted, or modified.

These definitions can blend together, and some summaries can be classified as either or both.  The overarching connection is that the summary size will ideally depend only on the approximation guarantee but not the size of the original data set, although in some cases logarithmic dependence is acceptable.

We focus on five types of coresets and sketches:
shape-fitting (Section 49.1),
density estimation (Section 49.2),
high-dimensional vectors (Section 49.3),
high-dimensional point sets / matrices (Section 49.4), and
clustering (Section 49.5).
There are many other types of coresets and sketches (e.g., for graphs~\cite{AGM12} or Fourier transforms~\cite{IKP14}) which we do not cover for space or because they are less geometric.

\Bnn{COMPUTATIONAL MODELS AND PRIMATIVES}

\noindent
Often the challenge is not simply to bound the size of a coreset or sketch as a function of the error tolerance, but to also do so efficiently and in a restricted model.  So before we discuss the specifics of the summaries, it will be useful to outline some basic computational models and techniques.

The most natural fit is a {\trmbitx streaming model} that allows limited space (e.g., the size of the coreset or sketch) and where the algorithm can only make a single scan over the data, that is one can read each data element once.
There are several other relevant models which are beyond the scope of this chapter to describe precisely.  Many of these consider settings where data is distributed across or streaming into different locations and it is useful to compress or maintain data as coresets and sketches at each location before communicating only these summaries to a central coordinator.
The {\trmbitx mergeable model} distills the core step of many of these distributed models to a single task: given two summaries $S_1$ and $S_2$ of disjoint data sets, with error bounds $\eps_1$ and $\eps_2$, the model requires a process to create a single summary $S$ of all of the data, of size $\max\{\textsf{size}(S_1), \textsf{size}(S_2)\}$, and with error bound $\eps = \max\{ \eps_1, \eps_2\}$.  Specific error bound definitions will vary widely, and will be discussed subsequently.
We will denote any such merge operation as $\oplus$, and a summary where these size and error constraints can be satisfied is called \emph{mergeable}~\cite{ACHPWY13}.

A more general {\trmbitx merge-reduce framework}~\cite{CM96,BS80} is also often used, including within the streaming model.  Here we may consider less sophisticated merge $\oplus$ operations, such as the union where the size of $S$ is $\textsf{size}(S_1) + \textsf{size}(S_2)$, and then a reduce operation to shrink the size of $S$, but resulting in an increased error, for instance as $\eps = \eps_1 + \eps_2$.
Combining these operations together into an efficient framework can obtain a summary of size $g$ (asymptotically, perhaps up to $\log$ factors) from a dataset of size $n$ as follows.
First arbitrarily divide the data into $n/g$ subsets, each of size $g$ (assume $n/g$ is a power of $2$, otherwise pad the data with dummy points).  Think of organizing these subsets in a binary tree.
Then in $\log(n/g)$ rounds until there is one remaining set, perform each of the next two steps.  First pair up all remaining sets, and merge each pair using an $\oplus$ operator.  Second, reduce each remaining set to be a summary of size $g$.
If the summary follows the mergeable model, the reduce step is unnecessary.

Even if the merge or reduce step requires some polynomial $m^c$ time to process $m$ data points, this is only applied to sets of size at most $2g$, hence the full runtime is dominated by the first round as $(n/g)\cdot(2g)^c = O(n \cdot g^{c-1})$.
The log factor increase in error (for that many merge-reduce steps) can be folded into the size $g$, or in many cases removed by delaying some reduce steps and careful bookkeeping~\cite{CM96}. 

In a streaming model this framework is applied by mapping data points to the $n/g$ subsets in the order they arrive, and then always completing as much of the merge-reduce process as possible given the data seen; e.g., scanning the binary tree over the initial subsets from left to right.
Another $\log (n/g)$ space factor is incurred for those many summaries which can be active at any given time.

\A{SHAPE FITTING}

\noindent
In this section we will discuss problems where given an input point set $P$, the goal is to find the best fitting shape from some class to $P$.  The two central problems in this area are the minimum (or smallest) enclosing ball, which has useful solutions in high dimensions, and the $\eps$-kernel coreset for directional width which approximates the convex hull but also can be transformed to solve many other problems.

\Bnn{GLOSSARY}

\begin{gllist}
\item {\index{minimum enclosing ball}\trmbitx Minimum enclosing ball (MEB):}\quad
Given a point set $P \subset \R^d$, it is the smallest ball $B$ which contains $P$.

\item {\index{eps-approximate minimum enclosing ball problem}\trmbitx $\eps$-Approximate minimum enclosing ball problem:}\quad
Given a point set $P \subset \R^d$, and a parameter $\eps >0$, the problem is to find a ball $B$ whose radius is no larger than $(1+\eps)$ times the radius of the MEB of $P$.

\item {\index{direction width}\trmbitx Ddirectional width:}\quad
Given a point set $P \subset \R^d$ and a unit vector $u \in \R^d$, then the \emph{directional width} of $P$ in direction $u$ is $\omega(P,u) = \max_{p \in P} \langle p, u \rangle - \min_{p \in P} \langle p, u \rangle$.

\item {\index{eps-kernel coreset}\trmbitx $\eps$-Kernel coreset:}\quad
An \emph{$\eps$-kernel coreset} of a point set $P \in \R^d$ is subset $Q \subset P$ so that for all unit vectors $u \in \R^d$, 
\[
0 \leq \omega(P,u) - \omega(Q,u) \leq \eps \omega(P,u).
\]

\item {\index{functional width}\trmbitx Functional width:}\quad
Given a set $\Eu{F} = \{f_1, \ldots, f_n\}$ of functions each from $\R^d$ to $\R$, the width at a point $x \in \R^d$ is defined $\omega_{\Eu{F}}(x) = \max_{f_i \in \Eu{F}} f_i(x) - \min_{f_i \in \Eu{F}} f_i(x)$.

\item {\index{eps-kernel for functional width}\trmbitx $\eps$-Kernel for functional width:}\quad
Given a set $\Eu{F} = \{f_1, \ldots, f_n\}$ of functions each from $\R^d$ to $\R$, an $\eps$-kernel coreset is a subset $\Eu{G} \subset \Eu{F}$ such that for all $x \in \R^d$ the functional width $\omega_{\Eu{G}}(x) \geq (1-\eps) \omega_{\Eu{F}}(x)$.

\item {\index{faithful measure}\trmbitx Faithful measure:}\quad
A measure $\mu$ is faithful if there exists a constant $c$, depending on $\mu$, such that for any point set $P \subset \R^d$ any $\eps$-kernel coreset $Q$ of $P$ is a coreset for $\mu$ with approximation parameter $c\eps$.

\item {\index{diameter}\trmbitx Diameter:}\quad
The diameter of a point set $P$ is $\max_{p,p' \in P} \|p-p'\|$.

\item {\index{width}\trmbitx Width:}\quad
The width of a point set $P$ is $\min_{u \in \R^d, \|u\|=1} \omega(P,u)$.

\item {\index{spherical shell}\trmbitx Spherical shell:}\quad
For a point $c \in \R^d$ and real numbers $0 \leq r \leq R$, it is the closed region $\sigma(c,r,R) = \{x \in \R^d \mid r \leq \|x-c\| \leq R\}$ between two concentric spheres of radius $r$ and $R$ centered at $c$.  Its \emph{width} is defined $R - r$.

\end{gllist}

\vspace{-.6pc}

\Bnn{SMALLEST ENCLOSING BALL CORESET}

\noindent
Given a point set $P \subset \R^d$ of size $n$, there exists a $\eps$-coreset for the smallest enclosing ball problem of size $\lceil 2/\eps \rceil$ that runs in time $O(nd/\eps + 1/\eps^5)$~\cite{BC03}.
Precisely, this finds a subset $S \subset P$ with smallest enclosing ball $B(S)$ described by center point $c$ and radius $r$; it holds that if $r$ is expanded to $(1+\eps)r$, then the ball with the same center would contain $P$.

The algorithm is very simple and iterative:
 At each step, maintain the center $c_i$ of the current set $S_i$, add to $S_i$ the point $p_i \in P$ furthest from $c_i$, and finally update $S_{i+1} = S_i \cup \{p_i\}$ and $c_{i+1}$ as the center of smallest enclosing ball of $S_{i+1}$.
Clarkson~\cite{Cla10} discusses the connection to the Frank-Wolfe~\cite{FW56} algorithm, and the generalizations towards several sparse optimization problems relevant for machine learning, for instance support vector machines~\cite{TKC05}, polytope distance~\cite{GJ09}, uncertain data~\cite{MSF14}, and general Riemannian manifolds~\cite{AN12}.  

These algorithms do not work in the streaming model, as they require $\Omega(1/\eps)$ passes over the data, but the runtime can be improved to $O((d/\eps + n/\eps^2)\log(n/\eps))$ with high probability~\cite{CHW10}.  Another approach~\cite{AS15} maintains a set of $O((1/\eps^3) \log(1/\eps))$ points in a stream that handles updates in $O((d/\eps^2) \log (1/\eps))$ time.  But it is not a coreset (a true proxy for $P$) since in order to handle updates, it needs to maintain these points as $O((1/\eps^2)\log(1/\eps))$ different groups.

\Bnnnr{EPSILON-KERNEL CORESET FOR WIDTH}

\noindent
Given point sets $P \subset \R^d$ of size $n$, an $\eps$-kernel coreset for directional width exists of size $O(1/\eps^{(d-1)/2})$~\cite{AHV04} and can be constructed in $O(n + 1/\eps^{d-(3/2)})$ time~\cite{Cha06,YAPV04}.  These algorithms are quite different than those for MEB, and the constants have heavy dependence on $d$ (in addition to it being in the exponent of $1/\eps$).  They first estimate the rough shape of the points so that they can be made fat (so width and diameter are $\Theta(1)$) through an affine transform that does not change which points form a coreset.  Then they carefully choose a small set of points in the extremal directions.

In the streaming model in $\R^d$, the $\eps$-kernel coreset can be computed using $O((1/\eps^{(d-1)/2}) \cdot \log(1/\eps))$ space with $O(1+ (1/\eps^{(d-3)/2})\log(1/\eps))$ update time, which can be amortized to $O(1)$ update time~\cite{ZZ08}.  In $\R^2$ this can be reduced to $O(1/\sqrt{\eps})$ space and $O(1)$ update time~\cite{AY07}.

Similar to $\eps$-kernels for directional width, given a set of $n$ $d$-variate linear functions $\Eu{F}$ and a parameter $\eps$, then an $\eps$-kernel for functional width can be computed of size $O(1/\eps^{d/2})$ in time $O(n + 1/\eps^{d-(1/2)})$~\cite{AHV04,Cha06}.

Many other measures can be shown to have $\eps$-approximate coresets by showing they are \emph{faithful}; this includes diameter, width, minimum enclosing cylinder, and minimum enclosing box.
Still other problems can be given $\eps$-approximate coresets by linearizing the inputs so they represent a set of $n$ linear functions in higher dimensions.  Most naturally this works for creating an $\eps$-kernel for the width of polynomial functions.  Similar linearization is possible for a slew of other shape-fitting problems including the minimum width spherical shell problem, overviewed nicely in a survey by Agarwal, Har-Peled and Varadarajan~\cite{AHV07}.

These coresets can be extended to handle a small number of outliers~\cite{HW04,AHY08} or uncertainty in the input~\cite{LPW14}.
A few approaches also extend to high dimensions, such as fitting a $k$-dimensional subspace~\cite{HV04,BHR16}.

\vspace{-0.6pc}
\A{DENSITY ESTIMATION}

\noindent
Here we consider a point set $P \subset \R^d$ which represents a discrete density function.  A coreset is then a subset $Q \subset P$ such that $Q$ represents a similar density function to $P$ under a restricted family of ways to measure the density on subsets of the domain, e.g., defined by a range space.

\vspace{-0.6pc}
\Bnn{GLOSSARY}

\begin{gllist}
\item {\index{range space}\trmbitx Range space:}\quad
A range space $(P,\Eu{A})$ consists of a ground set $P$ and a family of ranges $\Eu{R}$ of subsets from $P$.  In this chapter we consider ranges which are defined geometrically, for instance when $P$ is a point set and $\Eu{R}$ are all subsets defined by a ball, that is any subset of $P$ which coincides with $P \cap B$ for any ball $B$.

\item {\index{epsilon-net}\trmbitx $\eps$-Net:}\quad
Given a range space $(P,\Eu{R})$, it is a subset $Q \subset P$ so for any $R \in \Eu{R}$ such that $|R \cap P| \geq \eps |P|$, then $R \cap Q \neq \emptyset$.

\item {\index{epsilon-approximation}\trmbitx $\eps$-Approximation (or $\eps$-sample):}\quad
Given a range space $(P,\Eu{R})$, it is a subset $Q \subset P$ so for all $R \in \Eu{R}$ it implies $\left| \frac{|R \cap P|}{|P|} - \frac{|R \cap Q|}{|Q|} \right| \leq \eps$.

\item {\index{VC-dimension}\trmbitx VC-dimension:}\quad
For a range space $(P,\Eu{R})$ it is the size of the largest subset $Y \subset P$ such that for each subset $Z \subset Y$ it holds that $Z = Y \cap R$ for some $R \in \Eu{R}$.
\end{gllist}

\vspace{-.6pc}

\Bnnnr{RANDOM SAMPLING BOUNDS}

\noindent
Unlike the shape fitting coresets, these density estimate coresets can be constructed by simply selecting a large enough random sample of $P$.  The best such size bounds typically depend on VC-dimension $\nu$~\cite{VC71} (or shattering dimension $\sigma$), which for many geometrically defined ranges (e.g., by balls, halfspaces, rectangles) is $\Theta(d)$.
A random subset $Q \subset P$ of size $O((1/\eps^2)(\nu + \log(1/\delta))$~\cite{LLS01} is an $\eps$-approximation of any range space $(P, \Eu{R})$ with VC-dimension $\nu$, with probability at least $1-\delta$.
A subset $Q \subset P$ of size $O((\nu/\eps)\log(1/\eps\delta))$~\cite{HW87} is an $\eps$-net of any range space $(P, \Eu{R})$ with VC-dimension $\nu$, with probability at least $1-\delta$.

These bounds are of broad interest to learning theory, because they describe how many samples are sufficient to learn various sorts of classifiers.   In machine learning, it is typical to assume each data point $q \in Q$ is drawn iid from some unknown distribution, and since the above bounds have no dependence on $n$, we can replace $P$ by any probability distribution with domain $\R^d$.
Consider that each point in $Q$ has a value from $\{-,+\}$, and a separator range (e.g., a halfspace) should ideally have all $+$ points inside, and all $-$ points outside.
Then for an $\eps$-approximation $Q$ of a range space $(P, \Eu{A})$, the range $R \in \Eu{R}$ which misclassifies the fewest points on $Q$, misclassifies at most an $\eps$-fraction of points in $P$ more than the optimal separator does.
An $\eps$-net (which requires far fewer samples) can make the same claim as long as there exists a separator in $\Eu{A}$ that has zero misclassified points on $P$; it was recently shown~\cite{Han15} a weak coreset for this problem only requires $\Theta((1/\eps)(\nu + \log(1/\delta)))$ samples.

The typical $\eps$-approximation bound provides an additive error of $\eps$ in estimating $|R \cap P|/|P|$ with $|R \cap Q|/|Q|$.  One can achieve a stronger \emph{relative $(\rho,\eps)$-approximation} such that
\[
\max_{R \in \Eu{R}} \left| \frac{|R \cap P|}{|P|} - \frac{|R \cap Q|}{|Q|} \right| \leq \eps \max\left\{\rho, \frac{|R \cap P|}{|P|}\right\}.
\]
This requires $O((1/\rho\eps^2) (\nu \log (1/\rho) + \log(1/\delta)))$ samples~\cite{LLS01,HS11} to succeed with probability at least $1-\delta$.

\Bnn{DISCREPANCY-BASED RESULTS}

\noindent
Tighter bounds for density estimation coresets arise through discrepancy.
The basic idea is to build a coloring on the ground set $\chi : X \to \{-1,+1\}$ to minimize $\sum_{x \in R} \chi(x)$ over all ranges (the discrepancy).  Then we can plug this into the merge-reduce framework where merging takes the union and reducing discards the points colored $-1$.  Chazelle and Matou\v{s}ek~\cite{CM96} showed how slight modifications of the merge-reduce framework can remove extra log factors in the approximation.

Based on discrepancy results (see Chapters 14 and 48) we can achieve the following bounds.  These assume $d \geq 2$ is a fixed constant, and is absorbed in $O(\cdot)$ notation.
For any range space $(P, \Eu{R})$ with VC-dimension $\nu$ (a fixed constant) we can construct an $\eps$-approximation of size $g = O(1/\eps^{2 - \nu/(\nu+1)})$ in $O(n \cdot g^{w-1})$ time.
This is tight for range spaces $\Eu{H}_d$ defined by halfspaces in $\R^d$, where $\nu = d$.
For range spaces $\Eu{B}_d$ defined by balls in $\R^d$, where $\nu = d+1$ this can be improved slightly to $g = O(1/\eps^{2 - \nu/(\nu+1)} \sqrt{\log (1/\eps)})$; it is unknown if the $\log$ factor can be removed.
For range spaces $\Eu{T}_d$ defined by axis-aligned rectangles in $\R^d$, where $\nu = 2d$, this can be greatly improved to $g = O((1/\eps) \log^{d+ 1/2} (1/\eps))$ with the best lower bound as $g = \Omega((1/\eps) \log^{d - 1} (1/\eps))$ for $d \geq 2$~\cite{Lar14,MNT15}.
These colorings can be constructed adapting techniques from Bansal~\cite{Ban10,BS13}.  Various generalizations (typically following one of these patterns) can be found in books by Matou\v{s}ek~\cite{Mat10} and Chazelle~\cite{Cha01}.
Similar bounds exist in the streaming and mergeable models, adapting the merge-reduce framework~\cite{BCEG04,STZ04,ACHPWY13}.

Discrepancy based results also exist for constructing $\eps$-nets.  However often the improvement over the random sampling bounds are not as dramatic.  
For halfspaces in $\R^3$ and balls in $\R^2$ we can construct $\eps$-nets of size $O(1/\eps)$~\cite{MSW90,CV07,PR08}.
For axis-aligned rectangles and fat objects we can construct $\eps$-nets of size $O((1/\eps) \log \log (1/\eps))$~\cite{AES10}.
Pach and Tardos~\cite{PT13} then showed these results are tight, and that similar improvements cannot exist in higher dimensions.

\Bnn{GENERALIZATIONS}

\noindent
One can replace the set of ranges $\Eu{R}$ with a family of functions $\Eu{F}$ so that $f \in \Eu{F}$ has range $f : \R^d \to [0,1]$, or scaled to other ranges, including $[0,\infty)$.  For some $\Eu{F}$ we can interpret this as replacing a binary inclusion map $R : \R^d \to \{0,1\}$ for $R \in \Eu{R}$, with a continuous one $f : \R^d \to [0,1]$ for $f \in \Eu{F}$.
A family of functions $\Eu{F}$ is \emph{linked} to a range space $(P,\Eu{R})$ if for every value $\tau > 0$ and ever function $f \in \Eu{F}$, the points $\{p \in P \mid f(p) \geq \tau\} = R \cap P$ for some $R \in \Eu{R}$.
When $\Eu{F}$ is linked to $(P,\Eu{R})$, then an $\eps$-approximation $Q$ for $(P, \Eu{R})$ also $\eps$-approximates $(P,\Eu{F})$~\cite{JKPV11} (see also \cite{Har06,LS10} for similar statements) as
\[
\max_{f \in \Eu{F}} \left| \frac{\sum_{p \in P} f(p)}{|P|} - \frac{\sum_{q \in Q} f(q)}{|Q|} \right| \leq \eps.
\]
One can also show $\eps$-net type results.  An $(\tau,\eps)$-net for $(P,\Eu{F})$ has for all $f \in \Eu{F}$ such that $\frac{\sum_{p \in P} (p)}{|P|} \geq \eps$, then there exists some $q \in Q$ such at $f(q) \geq \tau$.
Then an $(\eps-\tau)$-net $Q$ for $(P,\Eu{R})$ is an $(\tau,\eps)$-net for $(P,\Eu{F})$ if they are linked~\cite{PZ15}.

A concrete example is for centrally-symmetric shift-invariant kernels $\Eu{K}$ (e.g., Gaussians $K(x,p) = \exp(-\|x-p\|^2)$) then we can set $f_x(p) = K(x,p)$.  Then the above $\eps$-approximation corresponds with an approximate kernel density estimate~\cite{JKPV11}.
Surprisingly, there exist discrepancy-based $\eps$-approximation constructions that are smaller for many kernels (including Gaussians) than for the linked ball range space; for instance in $\R^2$ with $|Q| = O((1/\eps) \sqrt{\log(1/\eps)})$~\cite{Phi13}.

One can also consider the minimum cost from a set $\{f_1, \ldots, f_k\} \subset \Eu{F}$ of functions~\cite{LS10}, then the size of the coreset often only increases by a factor $k$.  This setting will, for instance, be important for $k$-means clustering when $f(p) = \|x-p\|^2$ for some center $x \in \R^d$~\cite{FL11}.  And it can be generalized to robust error functions~\cite{FS12} and Gaussian mixture models~\cite{FFK11}.

\Bnn{QUANTILES SKETCH}

\noindent
Define the \emph{rank} of $v$ for set $X \in \R$ as $\textsf{rank}(X,v) = |\{x \in X \mid x \leq v\}|$.  A quantiles sketch $S$ over a data set $X$ of size $n$ allows for queries such that $|S(v) - \textsf{rank}(X,v)| \leq \eps n$ for all $v \in \R$.
This is equivalent to an $\eps$-approximation of a one-dimensional range space $(X,\Eu{I})$ where $\Eu{I}$ is defined by half-open intervals of the form $(-\infty,a]$.

A $\eps$-approximation coreset of size $1/\eps$ can be found by sorting $X$ and taking evenly spaced points in that sorted ordering.  Streaming sketches are also known; most famously the Greenwald-Khanna sketch~\cite{GK01} which takes $O((1/\eps) \log (\eps n))$ space, where $X$ is size $n$.  Recently, combining this sketch with others~\cite{ACHPWY13,FO15}, Karnin, Lang, and Liberty~\cite{KLL16} provided new sketches which require $O((1/\eps) \log \log(1/\eps))$ space in the streaming model and $O((1/\eps) \log^2 \log(1/\eps))$ space in the mergeable model.  

\A{HIGH DIMENSIONAL VECTORS}

\noindent
In this section we will consider high dimensional vectors $v = (v_1, v_2, \ldots, v_d)$.
When each $v_i$ is a positive integer, we can imagine these as the counts of a labeled set (the $d$ dimensions);  a subset of the set elements or the labels is a coreset approximating  the relative frequencies.
Even more generally, a sketch will compactly represent another vector $u$ which behaves similarly to $v$ under various norms.

\Bnn{GLOSSARY}

\begin{gllist}
\item {\index{ell-p norm}\trmbitx $\ell_p$-Norm:}\quad
For a vector $v \in \R^d$ the $\ell_p$ norm, for $p \in [1,\infty)$, is defined $\|v\|_p = (\sum_{i=1}^d |v_i|^p)^{1/p}$; if clear we use $\|v\| = \|v\|_2$.  For $p=0$ define $\|v\|_0 = |\{i \mid v_i \neq 0\}|$, the number of nonzero coordinates, and for $p=\infty$ define $\|v\|_\infty = \max_{i=1}^d |v_i|$.

\item {\index{k-sparse}\trmbitx $k$-Sparse:}\quad
A vector is $k$-sparse if $\|v\|_0 \leq k$.

\item {\index{additive ell_p/ell_q approximation}\trmbitx Additive $\ell_p / \ell_q$ approximation:}\quad
A vector $v$ has an additive  $\eps$-($\ell_p/\ell_q)$ approximation with vector $u$ if
$
\|v-u\|_p \leq \eps \|v\|_q.
$

\item {\index{k-sparse ell_p/ell_q approximation}\trmbitx $k$-Sparse $\ell_p / \ell_q$ approximation:}\quad
A vector $v$ has a $k$-sparse $\eps$-($\ell_p/\ell_q)$ approximation with vector $u$ if $u$ is $k$-sparse and
$
\|v-u\|_p \leq \eps \|v-u\|_q.
$

\item {\index{frequency count}\trmbitx Frequency count:}\quad
For a vector $v = (v_1, v_2, \ldots v_d)$  the value $v_i$ is called the $i$th frequency count of $v$.

\item {\index{frequency moment}\trmbitx Frequency moment:}\quad
For a vector $v = (v_1, v_2, \ldots v_d)$  the value $\|v\|_p$ is called the $p$th frequency moment of $v$.
\end{gllist}
\vspace{-0.6pc}

\Bnn{FREQUENCY APPROXIMATION}

\noindent
There are several types of coresets and sketches for frequency counts.
Derived by $\eps$-approximation and $\eps$-net bounds, we can create the following coresets over dimensions.  Assume $v$ has positive integer coordinates, and each coordinate's count $v_i$ represents $v_i$ distinct objects.  Then let $S$ be a random sample of size $k$ of these objects and $u(S)$ be an approximate vector defined so $u(S)_i = (\|v\|_1/k) \cdot |\{ s \in S \mid s = i\}|$.  Then with $k = O((1/\eps^2) \log(1/\delta))$ we have $\|v - u(S)\|_\infty \leq \eps \|v\|_1$ (an additive $\eps$-$(\ell_\infty/\ell_1)$ approximation) with probability at least $1-\delta$.  Moreover, if $k = O((1/\eps) \log(1/\eps \delta))$ then for all $i$ such that $v_i \geq \eps \|v\|_1$, then $u(S)_i \neq 0$, and we can then measure the true count to attain a weighted coreset which is again an additive $\eps$-$(\ell_\infty / \ell_1)$ approximation.  And in fact, there can be at most $1/\eps$ dimensions $i$ with $v_i \geq \eps \|v\|_1$, so there always exists a weighted coreset of size $1/\eps$.

Such a weighted coreset for additive $\eps$-($\ell_\infty/\ell_1)$ approximations that is $(1/\eps)$-sparse can be found deterministically in the streaming model via the Misra-Gries sketch~\cite{mg-fre-82} (or other variants~\cite{metwally06,DLM02,KSP03}).  This approach keeps $1/\eps$ counters with associated labels.  For a new item, if it matches a label, the counter is incremented, else if any counter is $0$ it takes over that counter/label, and otherwise, (perhaps unintuitively) all counters are decremented.

The count-min sketch~\cite{CM05} also provides an additive $\eps$-($\ell_\infty / \ell_1)$ approximation with space $O((1/\eps) \log (1/\delta))$ and is successful with probability $1-\delta$.
A count-sketch~\cite{CCM02} provides an additive $\eps$-($\ell_\infty / \ell_2)$ approximation with space $O((1/\eps^2) \log (1/\delta))$, and is successful with probability at least $1-\delta$.
Both of these linear sketches operate by using $O(\log 1/\delta)$ hash functions, each mapping $[d]$ to one of $O(1/\eps)$ or $O(1/\eps^2)$ counters.  The counters are incremented or decremented with the value $v_i$.  Then an estimate for $v_i$ can be recovered by examining all cells where $i$ hashes; the effect of other dimensions which hash there can be shown bounded.

\smallskip\noindent{\trmbitx Frequency moments.}
Another common task is to approximate the frequency moments $\|v\|_p$.  For $p =1$, this is the count and can be done exactly in a stream.  The AMS Sketch~\cite{AMS99} maintains a sketch of size $O((1/\eps^2) \log (1/\delta))$ that can derive a value $\widehat F_2$ so that $| \|v\|_2 - \widehat F_2 | \leq \eps \|v\|_2$ with probability at least $1-\delta$.

The FM Sketch~\cite{flajolet85:_probab} (and its extensions~\cite{AMS99,DF03}) show how to create a sketch of size $O((1/\eps^2) \log (1/\delta))$ which can derive an estimate $\widehat F_0$ so that $| \|v\|_0 - \widehat F_0 | \leq \eps \|v\|_1$ with probability at least $1-\delta$.
This works when $v_i$ are positive counts, and those counts are incremented one at a time in a stream.
Usually sketches and coresets have implicit assumptions that a ``word'' can fit $\log n$ bits where the stream is of size $n$, and is sufficient for each counter.  Interestingly and in contrast, these $\ell_0$ sketches operate with bits, and only have a hidden $\log \log n$ factor for bits.

\smallskip\noindent{\trmbitx $k$-sparse tail approximation.}
Some sketches can achieve $k$-sparse approximations (which are akin to coresets of size $k$) and have stronger error bounds that depend only on the ``tail'' of the matrix; this is the class of $k$-sparse $\eps$-($\ell_p/\ell_q$) approximations.  See the survey by Gilbert and Indyk for more details~\cite{GI10}.

These bounds are typically achieved by increasing the sketch size by a factor $k$, and then the $k$-sparse vector is the top $k$ of those elements.  The main recurring argument is roughly as follows:
  If you maintain the top $1/\eps$ counters, then the largest counter not maintained is of size at most $\eps \|v\|$.  Similarly, if you first remove the top $k$ counters (a set $K = \{i_1, i_2, \ldots, i_k\} \subset [d]$, let their collective norm be $\|v_K\|$), then maintain $1/\eps$ more, the largest not-maintained counter is at most $\eps (\|v\| - \|v_K\|)$.
The goal is then to sketch a $k$-sparse vector which approximates $v_K$; for instance the Misra-Gries Sketch~\cite{mg-fre-82} and Count-Min sketch~\cite{CM05} achieve $k$-sparse $\eps$-($\ell_\infty/\ell_1$)-approximations with $O(k/\eps)$ counters, and the Count sketch~\cite{CCM02} achieves $k$-sparse $\eps$-($\ell_\infty/\ell_2$)-approximations with $O(k^2/\eps^2)$ counters~\cite{BCIS10}.

\A{HIGH DIMENSIONAL POINT SETS (MATRICES)}
\label{sec:subspace-matrix}

\noindent
Matrix sketching has gained a lot of interest due to its close connection to scalability issues in machine learning and data mining.  The goal is often to replace a matrix with a small space and low-rank approximation.
However, given a $n \times d$ matrix $A$, it can also be imagined as $n$ points each in $\R^d$, and the span of a rank-$k$ approximation is a $k$-dimensional subspace that approximately includes all of the points.

Many of the techniques build on approaches from vector sketching, and again, many of the sketches are naturally interpretable as weighted coresets.  Sometimes it is natural to represent the result as a reduced set of rows in a $\ell \times d$ matrix $B$.  Other times it is more natural to consider the dimensionality reduction problem where the goal is an $n \times c$ matrix, and sometimes you do both!  But since these problems are typically phrased in terms of matrices, the difference comes down to simply transposing the input matrix.
We will write all results as approximating an $n \times d$ matrix $A$ using fewer rows, for instance, with an $\ell \times d$ matrix $B$.

Notoriously this problem can be solved optimally using the numerical linear algebra technique, the \emph{singular value decomposition}, in $O(nd^2)$ time.  The challenges are then to compute this more efficiently in streaming and other related settings.

We will describe three basic approaches (row sampling, random projections, and iterative SVD variants), and then some extensions and applications~\cite{FT15}.   The first two approaches are mainly randomized, and we will describe results with constant probability, and for the most part these bounds can be made to succeed with any probability $1-\delta$ by increasing the size by a factor $\log(1/\delta)$.

\Bnn{GLOSSARY}

\begin{gllist}
\item {\index{matrix rank}\trmbitx Matrix rank:}\quad
The \emph{rank} of an $n \times d$ matrix $A$, denoted $\textsf{rank}(A)$, is the smallest $k$ such that all rows (or columns) lie in a $k$-dimensional subspace of $\R^d$ (or $\R^n$).

\item {\index{singular value decomposition}\trmbitx Singular value decomposition:}\quad
Given an $n \times d$ matrix $A$, the singular value decomposition is a product $U \Sigma V^T$ where $U$ and $V$ are orthogonal, and $\Sigma$ is diagonal.  $U$ is $n \times n$, and $V$ is $d \times d$, and $\Sigma = \textsf{diag}(\sigma_1, \sigma_2, \ldots, \sigma_{\min\{n,d\}})$ (padded with either $n-d$ rows or $d-n$ columns of all $0$s, so $\Sigma$ is $n \times d$) where $\sigma_1 \geq \sigma_2 \geq \ldots \geq \sigma_{\min\{n,d\}} \geq 0$, and $\sigma_i = 0$ for all $i > \textsf{rank}(A)$.

The $i$th column of $U$ (resp. column of $V$) is called the $i$th left (resp. right) \emph{singular vector}; and $\sigma_i$ is the $i$th \emph{singular value}.

\item {\index{spectral norm}\trmbitx Spectral norm:}\quad
The spectral norm of matrix $A$ is denoted $\|A\|_2 = \displaystyle{\max_{x\neq 0}} \|A x\|/\|x\|$.

\item {\index{Frobenius norm}\trmbitx Frobenius norm:}\quad
The Frobenius norm of a matrix $A$ is $\|A\|_F = \sqrt{\sum_{i=1}^n \|a_i\|^2}$ where $a_i$ is the $i$th row of $A$.

\item {\index{low rank approximation of a matrix}\trmbitx Low rank approximation of a matrix:}\quad
The best rank $k$ approximation of a matrix $A$ is denoted $[A]_k$.  Let $\Sigma_k$ be the matrix $\Sigma$ (the singular values from the SVD of $A$) where the singular values $\sigma_i$ are set to $0$ for $i > k$.  Then $[A]_k = U \Sigma_k V^T$.   Note we can also ignore the columns of $U$ and $V$ after $k$; these are implicitly set to $0$ by multiplication with $\sigma_i = 0$. 
The $n \times d$ matrix $[A]_k$ is optimal in that over all rank $k$ matrices $B$ it minimizes $\|A - B\|_2$ and $\|A - B\|_F$.

\item {\index{projection}\trmbitx Projection:}\quad
For a subspace $F \subset \R^d$ and point $x \in \R^d$, define the projection $\pi_F(x) = \arg \min_{y \in F} \|x-y\|$.  For a $n \times d$ matrix $A$, then $\pi_F(A)$ defines the $n \times d$ matrix where each row is individually projected on to $F$.
\end{gllist}

\vspace{-0.6pc}

\Bnn{ROW SUBSET SELECTION}

\noindent
The first approach towards these matrix sketches is to chose a careful subset of the rows (note: the literature in this area usually discusses selecting columns).
An early analysis of these techniques considered sampling $\ell = O((1/\eps^2) k \log k)$ rows proportional to their squared norm as $\ell \times d$ matrix $B$, and showed~\cite{FKV04,DFKVV04,drineas2006fast2} one could describe a rank-$k$ matrix $P = [\pi_B(A)]_k$ so that
\[
\|A - P\|_F^2 \leq \|A - [A]_k\|_F^2  + \eps \|A\|_F^2
\hspace{.15in} \textrm{ and }  \hspace{.15in}
\|A - P\|_2^2 \leq \|A - [A]_k\|_2^2 + \eps \|A\|_F^2.
\]
This result can be extended for sampling columns in addition to rows.

This bound was then improved by sampling proportional to the \emph{leverage scores};  If $U_k$ is the $n \times k$ matrix of the first $k$ left singular vectors of $A$, then the leverage score of row $i$ is $\|U_k(i)\|^2$, the norm of the $i$th row of $U_k$.  In this case $O((1/\eps^2) k \log k)$ rows achieve a relative error bound~\cite{DMM08}
\[
\|A - \pi_B(A)\|_F \leq (1+\eps) \|A - [A]_k\|_F^2.
\]
These relative error results can be extended to sample rows and columns, generating a so-called CUR decomposition of $A$.
Similar relative error bounds can be achieved through volume sampling~\cite{DV06}.
Computing these leverage scores exactly can be as slow as the SVD; instead
one can approximate the leverage scores~\cite{DMMW12,CLMMPS15}, for instance in a stream in $O((kd/\eps^2) \log^4 n)$ bits of space~\cite{DMMW12}.

Better algorithms exist outside the streaming model~\cite{FVR15}.  These can, for instance, achieve the strong relative error bounds with only $O(k/\eps)$ rows (and $O(k/\eps)$ columns) and only require time $O(\textsf{nnz}(A) \log n + n \textsf{poly}(\log n, k, 1/\eps))$ time where $\textsf{nnz}(A)$ is the number of nonzero entries in $A$~\cite{BDM14,BW14}.
Or Batson, Spielman and Srivastava~\cite{BSS09,Sri10,Naor12} showed that $O(d/\eps^2)$ reweighted rows are sufficient and necessary to achieve bounds as below in (\ref{eq:OSE}).

\Bnn{RANDOM PROJECTIONS}

\noindent
The second approach to matrix sketching is based on the Johnson-Lindenstrauss (JL) Lemma~\cite{JL84}, which says that projecting any vector $x$ (independent of its dimension, for which it will be useful here to denote as $n$) onto a random subspace $F$ of dimension $\ell = O(1/\eps^2)$ preserves, with constant probability, its norm up to $(1+\eps)$ relative error, after rescaling:
$(1-\eps) \|x\| \leq \sqrt{n/\ell} \|\pi_F(x)\| \leq (1+\eps) \|x\|$.
Follow up work has shown that the projection operator $\pi_F$ can be realized as an $\ell \times n$ matrix $S$ so that $(\sqrt{n/\ell}) \pi_F(x) = Sx$.
And in particular, we can fill the entries of $S$ with iid Gaussian random variables~\cite{DG03},
uniform $\{-1,+1\}$ or $\{-1,0,+1\}$ random variables~\cite{Ach03}, or
any subgaussian random variable~\cite{Mat08}, rescaled; see Chapter 9 for more details.
Alternatively, we can make $S$ all $0$s except for one uniform $\{-1,+1\}$ random variable in each column of $S$~\cite{CW13}.  This latter construction essentially ``hashes'' each element of $A$ to one of elements in $\pi_F(x)$ (see also a variant~\cite{NN13}); basically an extension of the count-sketch~\cite{CCM02}.

To apply these results to matrix sketching, we simply apply the sketch matrix $S$ to $A$ instead of just a single ``dimension" of $A$.  Then $B = SA$ is our resulting $\ell \times d$ matrix.  However, unlike in typical uses of the JL Lemma on a point set of size $m$, where it can be shown to preserve all distances using $\ell = O((1/\eps^2) \log m)$ target dimensions, we will strive to preserve the norm over all $d$ dimensions.  As such we use $\ell = O(d/\eps^2)$ for iid JL results~\cite{Sar06}, or $\ell = O(d^2/\eps^2)$ for hashing-based approaches~\cite{CW13,NN13}.  As was first observed by Sarlos~\cite{Sar06} this allows one to create an oblivious subspace embedding so that for \emph{all} $x \in \R^d$ guarantees
\begin{equation} \label{eq:OSE}
(1-\eps)\|Ax\|_2^2 \leq \|Bx\|_2^2 \leq (1+\eps)\|Ax\|_2^2.
\end{equation}
The obliviousness of this linear projection matrix $S$ (it is created independent of $A$) is very powerful.  It means this result can be not only performed in the update-only streaming model, but also one that allows merges, deletions, or arbitrary updates to an individual entry in a matrix.  Moreover, given a matrix $A$ with only $\textsf{nnz}(A)$ nonzero entries, it can be applied in roughly $O(\textsf{nnz}(A))$ time~\cite{CW13,NN13}.  It also implies bounds for matrix multiplication, and as we will discuss, linear regression.

\Bnn{FREQUENT DIRECTIONS}

\noindent
This third class of matrix sketching algorithms tries to more directly replicate those properties of the SVD, and can be deterministic.  So why not just use the SVD?  These methods, while approximate, are faster than SVD, and work in the streaming and mergeable models.

The Frequent Directions algorithm~\cite{Lib13,FD-journal} essentially processes each row (or $O(\ell)$ rows) of $A$ at a time, always maintaining the best rank-$\ell$ approximation as the sketch.  But this can suffer from data drift, so crucially after each such update, it also shrinks all squared singular values of $[B]_\ell$ by $s_{\ell}^2$; this ensures that the additive error is never more than $\eps \|A\|_F^2$, precisely as in the Misra-Gries~\cite{mg-fre-82} sketch for frequency approximation.
Setting $\ell = k+1/\eps$ and $\ell = k + k/\eps$, respectively, the following bounds have been shown~\cite{GP14,FD-journal} for any unit vector $x$:
\[
0 \leq \|Ax\|^2 - \|Bx\|^2 \leq \eps \|A - [A]_k\|_F^2
\hspace{.15in} \text{ and } \hspace{.15in}
\|A - \pi_{[B]_k}(A)\|_F^2 \leq (1+\eps) \|A - [A]_k\|_F^2.
\]
Operating in a batch to process $\Theta(\ell)$ rows at a time, this takes $O(nd \ell)$ time.
A similar approach by Feldman et.al.~\cite{FSS13} provides a more general bound, and will be discussed in the context of subspace clustering below.

\Bnn{LINEAR REGRESSION AND ROBUST VARIANTS}

\noindent
The regression problem takes as input again an $n \times d$ matrix $A$ and also an $n \times w$ matrix $T$ (most commonly $w=1$ so $T$ is a vector); the goal is to find the $d \times w$ matrix $X^* = \arg\min_X \|AX - T\|_F$.
One can create a coreset of $\ell$ rows (or weighted linear combination of rows): the $\ell \times d$ matrix $\widehat A$ and $\ell \times w$ matrix $\widehat T$ imply a matrix $\widehat X = \arg\min_X \|\widehat A X - \widehat T\|_2$ that satisfies
\[
(1-\eps) \|A X^* - T\|_F^2 \leq \|A \widehat X - T\|_F^2 \leq (1+\eps) \|A X^* - T\|_F^2.
\]
Using the random projection techniques described above, one can sketch $\widehat{A} = SA$ and $\widehat T = SA$ with $\ell = O(d^2/\eps^2)$ for hashing approaches or $\ell = O(d/\eps^2)$ for iid approaches.  Moreover, Sarlos~\cite{Sar06} observed that for the $w=1$ case, since only a single direction (the optimal one) is required to be preserved (see \emph{weak coresets} below), one can also use just $\ell = O(d^2/\eps)$ rows.
Using row-sampling, one can deterministically select $\ell = O(d/\eps^2)$ rows~\cite{BDM13}.
The above works also provide bounds for approximating the multiple-regression spectral norm $\|AX^* - T\|_2$.

Mainly considering the single-regression problem when $w=1$ (in this case spectral and Frobenius norms bounds are equivalent $p=2$ norms), there also exists bounds for approximating $\|A X - T\|_p$ for $p \in [1,\infty)$ using random projection approaches and row sampling~\cite{CDMMMW13,Woo14}.  The main idea is to replace iid Gaussian random variables which are $2$-stable with iid $p$-stable random variables.  These results are improved using max-norm stability results~\cite{WZ13} embedding into $\ell_\infty$, or for other robust error functions like the Huber loss~\cite{CW15,CW15b}.

\A{CLUSTERING}

\noindent
An assignment-based clustering of a data set $X \subset \R^d$ is defined by a set of $k$ centers $C \subset \R^d$ and a function $\phi_C : \R^d \to C$, so $\phi_C(x) = \arg \min_{c \in C} \|x-c\|$.  The function $\phi_C$ maps to the closest center in $C$, and it assigns each point $x \in X$ to a center and an associated cluster.
It will be useful to consider a weight $w : X \to \R^+$.
Then a clustering is evaluated by a cost function
 \[
\cost_p(X,w,C) = \sum_{x \in X} w(x) \cdot \|x-\phi_C(x)\|^p.
 \]
For uniform weights (i.e., $w(x) = 1/|X|$, which we assume as default), then we simply write $\cost_p(X,C)$.  We also define $\cost_\infty(X,C) = \max_{x \in X} \|x - \phi_C(x)\|$.
These techniques extend to when the centers of the clusters are not just points, but can also be higher-dimensional subspaces.

\Bnn{GLOSSARY}

\begin{gllist}
\item {\index{k-means / k-median / k-center clustering problem}\trmbitx $k$-Means / $k$-median / $k$-center clustering problem:}\quad
Given a set $X \subset \R^d$, find a point set $C$ of size $k$ that minimizes $\cost_2(X,C)$ (respectively, $\cost_1(X,C)$ and $\cost_\infty(X,C)$).

\item {\index{(k,eps)-coreset for k-means / k-median / k-center}\trmbitx $(k,\eps)$-Coreset for $k$-means / $k$-median / $k$-center:}\quad
Given a point set $X \subset \R^d$, then a subset $S \subset X$ is a $(k,\eps)$-coreset for $k$-means (respectively, $k$-median and $k$-center) if for all center sets $C$ of size $k$ and parameter $p=2$ (respectively $p=1$ and $p=\infty$) that
\[
(1-\eps) \cost_p(X,C) \leq \cost_p(S,C) \leq (1+\eps) \cost_p(X,C).
\]

\item {\index{projective distance}\trmbitx Projective distance:}\quad
Consider a set $C = (C_1, C_2, \ldots, C_k)$ of $k$ affine subspaces of dimension $j$ in $\R^d$, and a power $p \in [1,\infty)$.
Then for a point $x \in \R^d$ the projective distance is defined $\dist_p(C,x) = \min_{C_i \in C} \|x - \pi_{C_i}(x)\|^p$, recalling that $\pi_{C_i}(x) = \arg \min_{y \in C_i} \|x-y\|$.

\item {\index{projective (k,j,p)-clustering problem}\trmbitx Projective $(k,j,p)$-clustering problem:}\quad
Given a set $X \subset \R^d$, find a set $C$ of $k$ $j$-dimensional affine subspaces that minimizes $\cost_p(X,C) = \sum_{x \in X} \dist_p(C,x)$.

\item {\index{(k,j,eps)-coreset for projective (k,j,p)-clustering}\trmbitx $(k,j,\eps)$-Coreset for projective $(k,j,p)$-clustering:}\quad
Given a point set $X \subset \R^d$, then a subset $S \subset X$, a weight function $w : S \to \R^+$, and a constant $\gamma$, is a $(k,j,\eps)$-coreset for projective $(k,j,p)$-clustering if for all $j$-dimensional center sets $C$ of size $k$ that
\[
(1-\eps) \cost_p(X,C) \leq \cost_p(X,w,C) + \gamma \leq (1+\eps) \cost_p(X,C).
\]
In many cases the constant $\gamma$ may not be needed; it will be $0$ unless stated.

\item {\index{strong coreset}\trmbitx Strong coreset:}\quad Given a point set $X \subset \R^d$, it is a subset $S \subset X$ that approximates the distance to any $k$-tuple of $j$-flats up to a multiplicative $(1+\eps)$ factor.

\item {\index{weak coreset}\trmbitx Weak coreset:}\quad
Given a point set $X \subset \R^d$, it is a subset $S \subset X$ such that the cost of the optimal solution (or one close to the optimal solution) of $(k,j)$-clustering on $S$, approximates the optimal solution on $X$ up to a $(1+\eps)$ factor.  So a strong coreset is also a weak coreset, but not the other way around.

\end{gllist}

\vspace{-0.6pc}

\Bnnnr{$k$-MEANS AND $k$-MEDIAN CLUSTERING CORESETS}
\label{sec:clustering}

\noindent
$(k,\eps)$-Coresets for $k$-means and for $k$-median are closely related.  The best bounds on the size of these coresets are independent of $n$ and sometimes also $d$, the number and dimension of points in the original point set $X$.
Feldman and Langberg~\cite{FL11} showed how to construct a strong $(k,\eps)$-coreset for $k$-median and $k$-means clustering of size $O(dk/\eps^2)$.
They also show how to construct a weak $(k,\eps)$-coreset~\cite{FMS07} of size $O(k \log(1/\eps)/\eps^3)$.
These bounds can generalize for any $\cost_p$ for $p \geq 1$.  However, note that for any fixed $X$ and $C$ that $\cost_p(X,C) > \cost_{p'}(X,C)$ for $p > p'$, hence these bounds are not meaningful for the $p = \infty$ special case associated with the $k$-center problem.
Using the merge-reduce framework, the weak coreset constructions work in the streaming model with $O((k/\eps^3) \log(1/\eps) \log^4 n)$ space.

Interestingly, in contrast to earlier work on these problems~\cite{HK07,HM04,Che06} which applied various forms of geometric discretization of $\R^d$, the above results make an explicit connection with VC-dimension-type results and density approximation~\cite{LS10,VX12}.  The idea is each point $x \in X$ is associated with a function $f_x(\cdot) = \cost_p(x,\cdot)$, and the total cost $\cost_p(X,\cdot)$ is a sum of these.
Then the mapping of these functions onto $k$ centers results in a generalized notion of dimension, similar to the VC-dimension of a dual range space, with dimension $O(kd)$, and then standard sampling arguments can be applied.

\Bnn{$k$-CENTER CLUSTERING CORESETS}

\noindent
The $k$-center clustering problem is harder than the $k$-means and $k$-median ones.  It is NP-hard to find a set of centers $\tilde C$ such that $\cost_\infty(X,\tilde C) \leq (2-\eta) \cost_\infty(X,C^*)$ where $C^*$ is the optimal center set and for any $\eta >0$~\cite{Hoc97}.
Yet, famously the Gonzalez algorithm~\cite{Gon85}, which always greedily chooses the point furthest away from any of the points already chosen, finds a set $\widehat C$ of size $k$, so $\cost_\infty(X,\widehat C) \leq 2 \cdot \cost_\infty(X,C^*)$.  This set $\widehat C$, plus the furthest point from any of these points (i.e., run the algorithm for $k+1$ steps instead of $k$) is a $(k,1)$ coreset (yielding the above stated $2$ approximation) of size $k+1$.
In a streaming setting, McCutchen and Khuller~\cite{MK08} describe a $O(k \log k \cdot (1/\eps) \log(1/\eps))$ space algorithm that provides $(2+\eps)$ approximation for the $k$-center clustering problem, and although not stated, can be interpreted as a streaming $(k,1+\eps)$-coreset for $k$-center clustering.

To get a $(k,\eps)$-coreset, in low dimensions, one can use the result of the Gonzalez algorithm to define a grid of size $O(k/\eps^d)$, keeping one point from each grid cell as a coreset of the same size~\cite{AP02}, in time $O(n + k/\eps^d)$~\cite{Har04a}.  In high dimensions one can run $O(k^{O(k/\eps)})$ sequences of $k$ parallel MEB algorithms to find a $k$-center coreset of size $O(k/\eps)$ in $O(dnk^{O(k/\eps)})$ time~\cite{BC03a}.

\Bnn{PROJECTIVE CLUSTERING CORESETS}
\label{sec:projection-clustering}

\noindent
Projective clustering seeks to find a set of $k$ subspaces of dimension $j$ which approximate a large, high-dimensional data set.  This can be seen as the combination of the subspace (matrix) approximations and clustering coresets.

Perhaps surprisingly, not all shape-fitting problems admit coresets -- and in particular subspace clustering ones pose a problem.  Har-Peled showed~\cite{Har04b} that no coreset exists for the 2-line-center clustering problem of size sublinear in the dataset.
This result can be interpreted so that for $j=1$ (and extended to $j=2)$, $k=2$, and $d=3$ then there is no coreset for projective $(k,j)$-center clustering problem sublinear in $n$.
Moreover a result of Meggido and Tamir~\cite{MT83} can be interpreted to say for $j \geq 2$ and $k > \log n$, the solution cannot be approximated in polynomial time, for any approximation factor, unless $P = NP$.

This motivates the study of bicriteria approximations, where the solution to the projective $(j,k,p)$-clustering problem can be approximated using a solution for larger values of $j$ and/or $k$.
Feldman and Langberg~\cite{FL11} describe a strong coreset for projective $(j,k)$-clustering of size  $O(djk/\eps^2)$ or weak coreset of size $O(kj^2 \log(1/\eps)/\eps^3)$, which approximated $k$ subspaces of dimension $j$ using $O(k \log n)$ subspaces of dimensions $j$.
This technique yields stronger bounds in the $j=0$ and $p=\infty$ case (the $k$-center clustering problem) where a set of $O(k \log n)$ cluster centers can be shown to achieve error no more than the optimal set of $k$ centers: a $(k,0)$-coreset for $k$-center clustering with an extra $O(\log n)$ factor in the number of centers.
Other tradeoffs are also described in their paper were the size or approximation factor varies as the required number of subspaces changes.
These approaches work in the streaming model with an extra factor $\log^4 n$ in space.

Feldman, Schmidt, and Sohler~\cite{FSS13} consider the specific case of $\cost_2$ and crucially make use of a nonzero $\gamma$ value in the definition of a $(k,j,\eps)$-coreset for projective $(k,j,2)$-clustering.
They show strong coresets
of size $O(j/\eps)$ for $k=1$ (subspace approximation),
of size $O(k^2/\eps^4)$ for $j=0$ ($k$-means clustering),
of size $\textsf{poly}(2^k, \log n,1/\eps)$ if $j=1$ ($k$-lines clustering), and
under the assumption that the coordinates of all points are integers between $1$ and $n^{O(1)}$,
of size $\textsf{poly}(2^{kj},1/\eps)$ if $j,k>1$.
These results are improved slightly in efficiency~\cite{CEMMP15}, and 
these constructions also extend to the streaming model with extra $\log n$ factors in space.

\A{SOURCES AND RELATED MATERIAL}

\vspace{-1pc}
\Bnn{SURVEYS}
\begin{trivlist}
\item[]\cite{AHV07}: A slightly dated, but excellent survey on coresets in geometry.

\item[]\cite{HPbook}: Book on geometric approximation that covers many of the above topics, for instance Chapters 5 ($\eps$-approximations and $\eps$-nets), Chapter 19 (dimensionality reduction), and Chapter 23 ($\eps$-kernel coresets).

\item[]\cite{Muthu05}: On Streaming, including history, puzzles, applications, and sketching.

\item[]\cite{Cormode11}: Nice introduction to sketching and its variations.

\item[]\cite{GI10}: Survey on $k$-sparse $\eps$-$(\ell_p/\ell_q)$ approximations.

\item[]\cite{Mah11,Woo14}: Surveys of randomized algorithms for Matrix Sketching.

\end{trivlist}

\vspace{-0.6pc}
\Bnn{RELATED CHAPTERS}

\noindent Chapter \phantom{1}9: Low-distortion embeddings of finite metric spaces

\noindent Chapter 14: Geometric discrepancy theory and uniform distribution

\noindent Chapter 48: Epsilon-nets and epsilon-approximations

\vspace{-1pc}

\Refh

\small
\bibliography{chap49}
\end{document}